# SIMULATING LARGE-SCALE STRUCTURE FORMATION FOR BSI POWER SPECTRA

Volker Müller
*Astrophysical Institute Potsdam, Germany.*

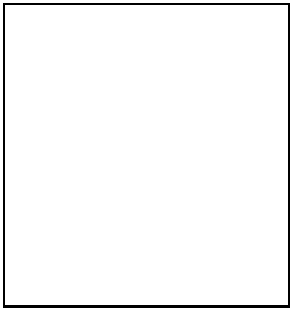

**Abstract**

A double inflationary model provides perturbation spectra with enhanced power at large scales (Broken Scale Invariant perturbations – BSI), leading to a promising scenario for the formation of cosmic structures. We describe a series of high-resolution PM simulations with a model for the thermodynamic evolution of baryons in which we are capable of identifying 'galaxy' halos with a reasonable mass spectrum and following the genesis of large and super-large scale structures. The power spectra and correlation functions of 'galaxies' are compared with reconstructed power spectra of the CfA catalogue and the correlation functions of the Las Campanas Deep Redshift Survey.

## 1 Double inflation scenario

We have discussed structure formation in a cosmological model with the dimensionless density parameter $\Omega = 1$, dominated by cold dark matter (CDM). However, alternatively to the standard CDM scenario, we used a primordial perturbation spectrum with a break at a characteristic scale. It is produced by the dominance of two scalar fields in the early cosmological evolution leading to a double inflation scenario. Original the idea goes back to Starobinsky [1], who considered $n$ inflaton fields producing an equivalent $n$-fold inflation. The logarithmic increase of the scale factor during such a scenario is given by the sum of the initial values of the different scalar field potentials. For a model with two fields we get a break in the primordial perturbation spectrum. We select parameters so that it lies at the comoving scale corresponding approximately to the initial radius of forming galaxy clusters. Then we get a typical scale which separates 'large' and 'small' cosmic structures.

The model used in our simulations was derived by Gottlöber, Müller, & Starobinsky [2]. A description of the power spectrum of primordial potential perturbations $P_\Phi(k)$ is given by the approximate formula (it does not describe the typical oscillations generated during the intermediate power law stage):

$$k^3 P_\Phi(k) = \begin{cases} 4.2 \times 10^{-6}[\log(k_s/k)]^{0.6} + 4.7 \times 10^{-6} & \text{for} \quad k < k_s \\ 9.4 \times 10^{-8} \log(k_f/k) & \text{for} \quad k > k_s \end{cases} \quad (1)$$

with $k_s = (2\pi/24)h^{-1}\text{Mpc}$ , $k_f = e^{56}h^{-1}\text{Mpc}$ . After folding this spectrum with the standard CDM transfer functions (using a dimensionless Hubble constant $h = 0.5$ besides $\Omega = 1$) we get the power spectrum shown by the dash-dotted line in Fig. 1 (left). For comparison we show the CDM curve as dotted line, it has more small scale power.

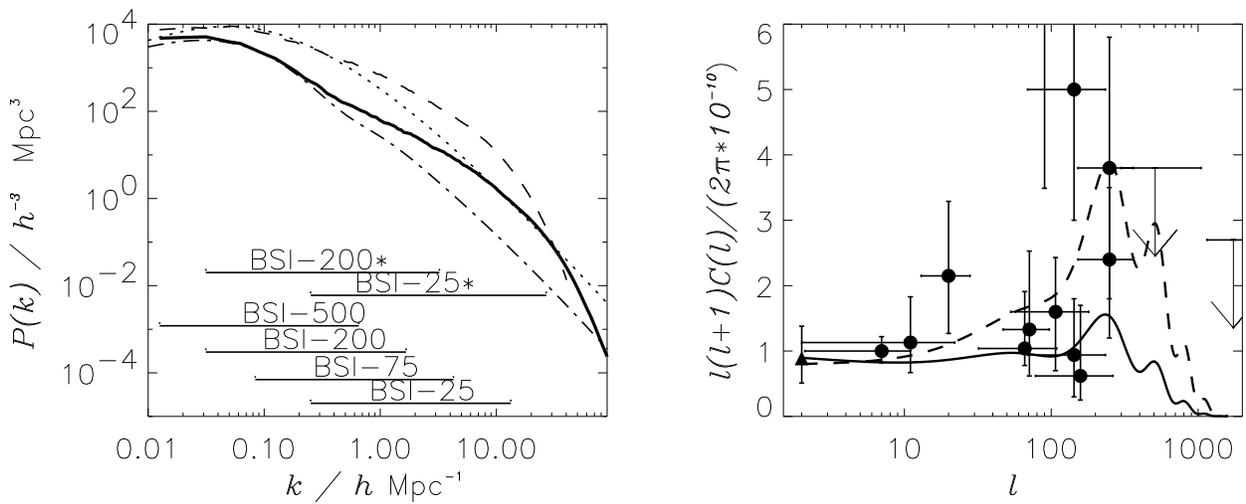

Figure 1: BSI and CDM power spectra (left) and inhomogeneities in the background radiation (right).

At decoupling the primordial potential perturbations impose imprints in the temperature fluctuations whose angular spectrum differs from the standard CDM prediction. Fig. 1 (right) shows the comparision of the BSI spectrum (solid line) and the CDM spectrum (dashed line) with a set of observations. As indicated the power spectra are normalised by the most reliable measurements at small multipoles (COBE-normalisation), but they differ in the amplitude of the Doppler peak at about $l \approx 200$.

## 2   N-body simulation and galaxy identification

At presence no theoretical description or simulation scheme exhaust all the various effects of structure formation in the universe. For testing our envisaged model we have to study both the large scale structure formation, i.e. the formation of cluster and superclusters of galaxies, and the formation of the different types of galaxies as the basic building blocks of all large cosmic structures.

Therefore we studied a collection of simulation boxes which range from 25 $h^{-1}$Mpc for having enough resolution at galactic scales, to 75 $h^{-1}$Mpc and 200 $h^{-1}$Mpc for study galaxy clustering, and to 500 $h^{-1}$Mpc for the super-large scale structure as studied in [3]. We used the PM code of Kates, Kotok, & Klypin [4], taking for BSI in the highest resolution $512^3$ cells and $256^3$ particles (simulation names with asterisk in Fig. 1) and a set of simulations with half the resolution for comparing BSI and CDM [5]. The range realised by the different simulations is indicated in Fig. 1. The solid line shows the combined non-linear spectrum of the dark matter after evolving from redshift $z = 25$ to the presence. The higher non-linear power of the corresponding CDM-simulations over a wide range of scales is shown by the dashed line.

During the simulation the code looks for shocks between triplets of originally neighboring particles, i.e. it looks for changes of the sign of the volume spanned by these particles. At shocks we attribute to the particles a 'temperature', which characterises the mean baryonic matter comoving with the dark matter (we suppose a cosmic abundance of 10 percent baryons). Subsequently we solve the cooling (and adiabatic heating) equations for these particles. This recipe leads to a reasonable account of the condensation of baryonic matter in galaxy halos and its clustering [5]. Since we do not alter the dynamics during and after shocks, dark and baryonic matter remains in the same pattern. Further we do not account of secondary shocks and of the heating by the intergalactic UV background. But we allow a certain fraction (85%) of reheating of the baryonic matter which became cold. This is believed to describe a phenomenological feedback mechanism provided by supernovae explosion in the forming galaxies. Finally it is essential that particles can get heated by neighboring high temperature particles.

Due to the nonlinear gravitational instability a large part of 'baryons' (up to 88%) gets 'cold' and transforms to star and star systems. Galaxies are identified with density maxima in the distribution of cold particles, which lie over a certain threshold (2 to 3 times the grid variance), but 'galaxy halos' are assembled by all particles within a search radius of 0.5 times the mean interparticle separation. The resulting mass distribution can be fitted by a Schechter function

$$\mathrm{d}n/\mathrm{d}\log M = n_*(M/M_*)^{-p}\exp(-M/M_*), \tag{2}$$

For BSI-25* we get $M_* = 10^{12} M_\odot$, $n_* = 2.6 \times 10^{-2} h^3 \mathrm{Mpc}^{-3}$ and $p = 0.8$. This is a quite reasonable density of $M_*$ galaxies (corresponding to the characteristic $L_*$ galaxies). Their spatial distribution can be used to evaluate the astrophysical effects of the BSI spectrum.

## 3 Power spectra and correlation function

Here we describe first the comparison of simulated power spectra with the power spectrum reconstructed from the CfA galaxy catalogue [6]. We impose redshift corrections by placing an arbitrary observer far outside the simulation box. The resulting power spectrum of BSI and CDM galaxies, identified in simulations with a box length of 200 $h^{-1}$Mpc is shown in Fig. 2 (left), where we select 'galaxies' with more than 10 particles for BSI and over 30 particles for CDM (the higher threshold is due to the more advanced clustering in the CDM model). The CDM model is strongly influenced by the redshift corrections which enhance the power near the maximum and suppress it at higher $k$ values (another cause for suppression of the power at high wave numbers is the subtraction of shot noise). Both spectra can fit the CfA power spectrum, in particular the BSI model leads to a very good fit of the data. The galaxies used in this simulation are biased with respect to the sea of all dark matter particles by an (approximately linear) bias factor of about 2, contrary to the CDM model which has $b \approx 1$.

The galaxy correlation function shown in Fig. 2 (right) supplements the study of the power spectrum. While in principle it has the same information content, the accuracy of its estimation is much higher in the region of nonlinear clustering. Here we compare 'galaxies' composed of a minimum number of 5 and 10 particles in the BSI and CDM simulations, respectively. The data from the Las Campanas Deep Redshift Survey are taken from [7]. Again, due to redshift corrections both curves are able to reproduce the observed galaxy clustering over a wide range of scales (in the contrary, the real space correlation function of CDM galaxies is much steeper than that of BSI galaxies). The lower clustering level of the data at lengths below 2 $h^{-1}$Mpc is probably due to the selection effects. A small difference becomes visible in the length range $(20-30)$ $h^{-1}$Mpc, where the higher clustering of the LCDRS galaxies may indicate an enhanced power, it is slightly better reproduced by the BSI model. The dash-dotted line in Fig. 2 (right) shows the best fit power law $\xi = (r/r_0)^{-1.6}$ with a correlation length of $r_0 = 5$ $h^{-1}$Mpc.

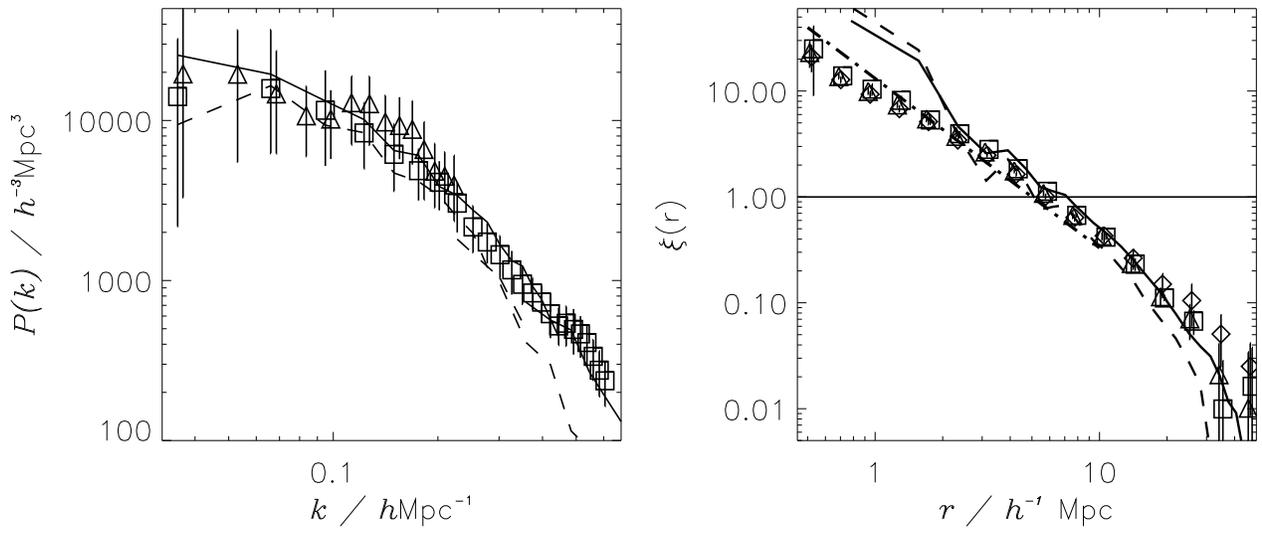

Figure 2: Galaxy power spectra and two-point correlation function.

Strong differences between the properties of galaxies in the two models become visible in the small scale velocity dispersion. Its one dimensional (line of sight) projection value lies at about 250 km/s in the BSI model and at about 600 km/s in the CDM model. This velocity dispersion can be estimated from the anisotropy of the two-point correlation function of redshift catalogues with respect to the line of sight and orthogonal to it. Further strong differences come from the observed mass distribution of galaxy clusters and the cluster-cluster correlation function ([8]). In the BSI model, the latter remains positive on scales up to at least 60 $h^{-1}$Mpc.

**Acknowledgements.** I am grateful to S. Gottlöber, R. Kates, J. Mücket, J. Retzlaff, and D. Tucker for a stimulating cooperation.